\documentclass[draftmode]{AEA}
\usepackage{amsmath}

\usepackage{graphicx}
\draftSpacing{1.5}

\def\monthname{\ifcase\month\or
  January\or February\or March\or April\or May\or June\or July\or
  August\or September\or October\or November\or December\fi}

\renewcommand{\thepage}{[\arabic{page}]}%
\numberwithin{equation}{section}

\newcommand{\tick}{\surd}

\begin{document}

\title{Ensemble Methods for Causal Effects in Panel Data Settings}
\shortTitle{Robustness}
\author{Susan Athey,  Mohsen Bayati, Guido Imbens, and Zhaonan Qu\thanks{Athey: Graduate School of Business, Stanford University, and NBER, 655 Knight Way, Stanford, CA 94305, athey@stanford.edu. 
Bayati: Graduate School of Business, Stanford University, 655 Knight Way, Stanford, CA 94305, bayati@stanford.edu. 
Imbens: Graduate School of Business, Stanford University, and NBER, 655 Knight Way, Stanford, CA 94305, imbens@stanford.edu. 
Qu: Department of Economics, Stanford University, Stanford, CA 94305, zhaonanq@stanford.edu. }}
\begin{abstract}
This paper studies a panel data setting where the goal is to estimate causal effects of an intervention by predicting the counterfactual
values of outcomes for treated units, had they not received the treatment.  Several approaches have been proposed for this
problem, including regression methods, synthetic control methods and matrix completion methods.  This paper considers an ensemble
approach, and shows that it performs better than any of the individual methods in several economic datasets.  Matrix completion methods are often given the most weight by the ensemble, but this clearly depends on the setting.  We argue that ensemble methods present a fruitful direction for further research in the causal panel data setting.
\end{abstract}
\pubMonth{Month}
\date{\today}
\pubYear{2019}
\pubVolume{Vol}
\pubIssue{Issue}
\maketitle

In many prediction problems researchers have found that combinations  of prediction methods (``ensembles'') perform better than individual methods (Hal Varian, 2014; James Surowiecki, 2005; Zhi-Hua Zhou, 1983).
A simple example is random forests, which combines predictions from many regression trees.
A striking, and substantially more complex, example is the Netflix Prize competion (e.g., Robert Bell, Yehuda Koren and Chris Volinsky (2010)) where the winning entry combined predictions using a wide variety of conceptually very different models. In macro-economic forecasting researchers have often found that averaging predictions from different models leads to more accurate forecasts (e.g., James Stock and Mark Watson, 2006).

In this paper we apply these ideas to synthetic control type problems in panel data settings (Alberto Abadie, Alexis Diamond and Jens Hainmueller, 2010,2015). In this setting a number of conceptually quite different methods have been developed, with some assuming  correlations between units that are stable over time, others assuming stable time series patterns common to all units, and others using factor models. With data on state level GDP for 270  quarters, we focus on three basic approaches to predicting missing values, one from each of these strands of the literature. 
Rather than try to test the different models against each other and find a true model (Quang Vuong, 1989), we focus on combining predictions based on each of the separate models using ensemble methods.
For the ensemble predictor we focus on a weighted average of the three individual methods, with non-negative weights determined through out-of-sample cross-validation. Alternatives to this stacked regression method (Leo Breiman, 1996) include   ridge regression (Mark van der Laan, Eric Polley, and Alan Hubbard, 2007), and Bayesian model averaging 
(Jennifer Hoeting, David Madigan, Adrian Raftery, and Chris  Volinsky, 1999).

Using a range of outcomes, GDP, log GDP, and GDP growth rates, and a range of values for the number of time periods, from $T=10$ to $T=270$, we find that no individual method dominates. Using the ensemble method with a convex combination of the individual methods we do substantially better, and typically as well as or better than the best individual method, in line with the general ensemble literature. These results show that ensemble methods are a practical and effective method for the type of data configurations typically encountered in empirical work in economics, and that these methods deserve more use in practice then they have received.

Many questions remain concerning their use in practice. Instead of linear regression to determine the weights of the individual predictors one can use more sophisticated methods, e.g., random forests, or neural nets. Choices regarding the cross-validation also matter. Finally, there are questions regarding inference  for the ensemble methods.

\section{A Simple Example}

Here we briefly discuss a very simple example to set the stage for the subsequent discussions. Suppose we have $N$ observations on an outcome and two covariates, $(Y_i,X_{i1},X_{i2})$, for $i=1,\ldots,N$. Let us consider two simple models, one that relates $Y_i$ to $X_{i1}$ through a linear model, and one that relates $Y_i$ to $X_{i2}$ through a linear model. Let the estimated regression functions be $\hat Y_{1i}=\hat\beta_{1,0}+\hat\beta_{1,1} X_{i1}$ and $\hat Y_{2i}=\hat\beta_{2,0}+\hat\beta_{2,2} X_{i2}$. 

We can combine these two models in different ways. 
One traditional approach would be to choose between the two models using statistical tests, as in Vuong (1989).
A second possibility  would be to build a more general model that nests both individual models, e.g., ${E}[Y_i|X_{i1},X_{i2}]=\beta_0+\beta_1X_{i1}+\beta_2 X_{i2}$. A third approach would be to assume that the true data generating process is a mixture of the two models, $Y_i=D_i(\beta_{1,0}+\beta_{1,1} X_{i1}+\varepsilon_{1,i})+(1-D_i)(\beta_{2,0}+\beta_{2,2} X_{i2}+\varepsilon_{2,i})$, with $D_i$ having a binomial distribution. Such a mixture model  nests both individual models, but in a different way. As a fourth approach, an implementation of an ensemble method is to predict future values for the outcome as a linear combination of the two individual predictors, $\hat Y_i=\theta_0+\theta_1 \hat Y_{1i}+\theta_2 \hat Y_{2i}$, with the weights $\theta_0$, $\theta_1$, and $\theta_2$ chosen to optimize the fit, possibly under the restriction that $\theta_0=0$ and $\theta_1+\theta_2=1$.

There are some insights that are present even in this simple case. Choosing between the models, based on statistical tests or out-of-sample fit, is not necessarily the optimal thing to do. Suppose that $X_{i1}$ and $X_{i2}$ are independent, and both contribute to the explanation of $Y_i$. In that case a model that uses both will generally outperform one that chooses one model or the other. Similarly a mixture model will not necessarily lead to the best predictions  if the true data generating process is $Y_i\sim {\cal N}(\beta_0+\beta_1X_{i1}+\beta_2X_{i2},\sigma^2)$, with $\beta_1,\beta_2\neq 0$. In this simple case either estimating the more general model with ${E}[Y_i|X_{i1},X_{i2}]=\beta_0+\beta_1X_{i1}+\beta_2 X_{i2}$, or using an ensemble method that looks for the optimal linear combination of $\hat Y_{i1}$ and $\hat Y_{i2}$ will find the optimal linear combination of the two predictors in large samples.

Why might ensemble methods be an attractive way to go, compared to, say building a more general model that nests those that underly the individual predictorss or to using a mixture model? One important one is computational, with ensemble methods  computationally feasible and attractive even in settings with many individual predictors and with many observations, possibly on a large number of variables, and mixture methods relying in the EM algorithm can be slow. Second, building general models that nest simpler ones can be challenging, as will be clear in the setting in this paper. It is useful to note that a mixture model would ultimately estimate the individual models only on the data that appear to arise from those models, whereas ensemble models are estimated on al the data. From that perspective an important aspect of the choice between mixture models and ensemble models is the implicit regularization.

\section{Methods for Synthetic Control Problems}

Suppose we are interested in estimating the causal effect of an intervention in a setting with panel data. We observe outcomes for $N$ units over $T$ time periods. Suppose that only a single unit is ever exposed to the treatment, and this unit is only exposed in the last period. (Although the ideas discussed in this paper extend to more general settings, e.g., staggered adoption as in Susan Athey and Guido Imbens (2018), we focus on this simple setting for expository reasons.) So, for unit $N$ in period $T$ we observe $Y_{NT}=Y_{NT}(1)$, and for all other unit/time pairs $(i,t)$ we observe $Y_{it}=Y_{it}(0)$. To estimate the causal effect $\tau=Y_{NT}(1)-Y_{NT}(0)$ we wish to impute the missing $Y_{NT}(0)$ based on the $NT-1$ observations on $Y_{it}(0)$. 
See Equation (\ref{een}) for the data configuration.
{\small
\begin{equation}\label{een}\mathbf{Y}(0) =\left(
\begin{array}{cccccc}
 \tick & \tick & \tick  & \dots & \tick& \tick\\
\tick  & \tick & \tick   & \dots & \tick& \tick  \\
\vdots   &  \vdots & \vdots &\ddots & \vdots  &\vdots \\
\tick  & \tick & \tick   & \dots & \tick& \tick  \\
\tick  & \tick & \tick   & \dots & \tick& ?  \\
\end{array}
\right).\end{equation}
}
Many conceptually quite distinct methods have been proposed in the literature for imputing $Y_{NT}(0)$.  Leading methods include $(i)$ methods based on unconfoundedness, 
$(ii)$ synthetic control methods (Alberto Abadie, Alexis Diamond and Jens Hainmueller, 2010,2015), 
and $(iii)$ matrix completion methods that fit factor models to $\mathbf{Y}$ (e.g., Jushan Bai and Serena Ng, 2007;  Athey, Mohsen Bayati, Nikolay Doudchenko, Imbens, and Khashayar Khosravi, 2018).  Choosing between the different approaches is not always straightforward.The relative magnitude of $N$ and $T$, and the correlation patterns matter substantially. In this paper
we show that the relative performace of the three methods varies substantially by the magnitude of $T$ (for a fixed $N$), and by the nature of the variable.
 It may therefore be unappealling to choose a priori one of the different approaches. However, this is not necessary. One alternative is to use ensemble or model averaging methods that combine the different approaches. Here we describe concretely how such model averaging would work in settings with conceptually quite different models, and we see that the ensemble methods systematically perform well relative to the individual methods. Next we briefly describe the three individual methods.

\subsection{Synthetic Control Methods or Vertical Regression (VR)}

First we describe  synthetic control methods or vertical regression, where the unit of observation is a column in $\mathbf{Y}$. See Abadie, Diamond and Hainmueller (2010, 2015) for more details. Here we focus on the regularized unconstrained weights, defined as
\[ \arg\min_{\omega_0,\omega} \Biggl\{\sum_{t=1}^{T-1}\left(Y_{Nt}-\omega_0-\sum_{i=1}^{N-1}\omega_i Y_{it}\right)^2\]
\[\hskip1cm +\lambda\left( \alpha ||\omega\|_1+\frac{1-\alpha}{2}\|\omega\|^2_2\right)\Biggr\}.\]
 The imputed value is then $Y^{\sc vt}_{NT}=\omega_0+\sum_{i=1}^{N-1} \omega_i Y_{iT}$. The standard SC estimator imposes the additional restriction that $\omega_0=0$, $\sum_{i=1}^{N-1}\omega_i=1$, and $\omega_i\geq 0$, and does not use the elastic net penalty term which we include to deal with settings where $N$ is large relative to $T$.
 The penalty parameters $\lambda$ and $\alpha$ are chosen through crossvalidation using different time periods.

\subsection{Unconfoundedness or Horizontal Regression (HZ)}

The second method we consider is a regression based on an unconfoundedness assumption (e.g., Imbens and Donald Rubin, 2015), that the treatment assignment is independent of the potential outcomes. The estimator we consider is based on a regression of the final period outcome on the lagged outcomes:
\[ \arg\min_{\beta_0,\beta}\Biggl\{\sum_{i=1}^{N-1} \left(Y_{iT}-\beta_0-\sum_{t=1}^{T-1}\beta_t Y_{it}\right)^2\]
\[\hskip1cm +\lambda\left( \alpha ||\beta\|_1+\frac{1-\alpha}{2}\|\beta\|^2_2\right)\Biggr\}.\]
The penalty parameters here are again chosen by crossvalidation, this time using different units.
 The imputed value is then $Y^{\sc hz}_{NT}=\beta_0+\sum_{t=1}^{T-1} \beta_t Y_{Nt}$.  Note that this estimator is based on the same algorithm as the vertical regression estimator after we transpose the matrix ${\bf Y}$.

\subsection{Matrix Completion (MC) Methods}

The third method we consider is based on matrix completion methods. 
See Athey, Bayati, Doudchenko, Imbens, and Khosravi (2018) for details.
We solve
\[ \min_{\mathbf{L},\alpha,\beta}
\sum_{(i,t)\neq (N,T)} \left( Y_{it}-\alpha_i-\beta_t-L_{it}\right)^2\]
\[\hskip2cm+\lambda \|\mathbf{L}\|_*,\]
where $\|\mathbf{L}\|_*$ is the nuclear norm, equal to the sum of the singular values in the singular value decomposition of $\mathbf{L}$. This leads to a factor model where $L_{it}=\sum_{r=1}^R A_{ir}B_{tr}$ with the rank $R$ determined through the nuclear norm penalization. The imputed value is then $\hat Y_{NT}^{\sc mc}=L_{NT}+\alpha_N+\beta_T$. The penalty parameter is choosen again through crossvalidation. Note that we include time and unit fixed effects in the model. These could be subsumed in the matrix ${\bf L}$, but in practice including them without penalization leads to superior out of sample fit, the same way the intercept in LASSO is not penalized.

\section{Ensemble Methods}

Here we discuss two method for combining the three individual methods for predicting the counterfactual values for unit $N$ in period $T$ that differ in how the weights are chosen. In both cases we focus on stacked regression (Leo Breiman, 1996). 

\subsection{Vertical Crossvalidation (VC)}

For the  VC method we first go through all the control units $i=1,\ldots,N$. Putting aside their final period outcome $Y_{iT}$ as well as the observations on the $N$-th unit, we use the three methods, VR, HZ, and MC to obtain three estimates,  $\hat Y_{iT}^{\sc vt}$, 
$\hat Y_{iT}^{\sc hz}$, and $\hat Y_{iT}^{\sc mc}$.  Next
 we estimate the weights as
\[\min_{\theta\geq 0}\sum_{i=1}^{N-1}\left( Y_{iT}-\theta_{\sc vt} \hat Y_{iT}^{\sc vt}-
\theta_{\sc hz} \hat Y_{iT}^{\sc hz}-
\theta_{\sc mc} \hat Y_{iT}^{\sc mc}\right)^2,\]
with the restriction  $\theta_{\sc vt}+\theta_{\sc hz}+\theta_{\sc mc}=1$. 
Then the ensemble estimator for the counterfactual value $Y_{NT}(0)$ is
\[ \hat Y_{NT}^{\sc ens,vc}=
\theta_{\sc vt} \hat Y_{NT}^{\sc vt}+
\theta_{\sc hz} \hat Y_{NT}^{\sc hz}+
\theta_{\sc mc} \hat Y_{NT}^{\sc mc}.\]

\subsection{Horizontal Crossvalidation (HC)}

 For the HC  method, we first go through the last $S$ pre-treatment periods, $s=1,\ldots,S$. Putting aside  outcome $Y_{NT-s}$, we use the three methods, VR, HZ, and MC to obtain three estimates,  $\hat Y_{NT-s}^{\sc vt}$, 
$\hat Y_{NT-s}^{\sc hz}$, and $\hat Y_{NT-s}^{\sc mc}$.  Now
 we estimate the weights as
\[\min_{\theta\geq 0}\sum_{s=1}^{S}\left( Y_{NT-s}-\theta_{\sc vt} \hat Y_{NT-s}^{\sc vt}-
\theta_{\sc hz} \hat Y_{NT-s}^{\sc hz}\right.\]
\[\hskip2cm \left.-
\theta_{\sc mc} \hat Y_{NT-s}^{\sc mc}\right)^2,\]
again with the restriction  $\theta_{\sc vt}+\theta_{\sc hz}+\theta_{\sc mc}=1$. Note that this method is not symmetric with the vertical crossvalidation - we only use the last $S$ periods for the crossvalidation to avoid predicting the past with the future.

\section{An Application}

Here we evaluate the three individual methods and the ensemble method in a realistic setting where we know the truth so we can obtain a realistic assessment of the relative merits.

\subsection{Set Up}

We take as our  outcomes state GDP, log GDP and GDP growth rates. 
This gives us a wide range of  correlation patterns across time and states. 
We then carry out the following exercise. Fixing $N=51$ (for the fifty states plus DC), and for four different values of $T$, $T\in\{10,25,100,270\}$, we take the last $T$ periods in our sample.  The variation in the number of time periods puts varying degrees of emphasis on the quality of the penalization methods. For example, for the horizontal regression methods we estimate a linear regression with 50 observations and 270 regressors.
With a large number of time periods we also expect that methods that treat all time periods as equally informative may perform poorly.

{\small
 \begin{table} \begin{center}
\begin{tabular}{lccccccccc}
\multicolumn{9}{c}{\sc Table I: } \\
 &\multicolumn{5}{c}{Average Root-Mean-Squared-Error}& &\multicolumn{3}{c}{Weight in VC}\\ 
  &\multicolumn{5}{c}{}& &\multicolumn{3}{c}{ Ensemble}\\
$\#$ Periods & VR & HR & MC &Ens-VC & Ens-HC&\ \ & VR & HR & MC\\
\multicolumn{5}{l}{GDP}\\
10  & 0.48 & 1.37 & 0.75 & 0.44&0.48 & &0.47 & 0.17 & 0.36  \\
25  & 0.12 & 0.37 & 0.13 & 0.12 &0.10&& 0.19 & 0.02 & 0.79\\
100  &0.16 & 0.15 & 0.04&  0.04&0.05& &0.00 & 0.00 & 1.00 \\
270  & 0.15 & 0.11 & 0.04 & 0.07&0.04&& 0.00 & 0.46 & 0.54 \\
&\\
\multicolumn{5}{l}{Log GDP}\\
10 & 0.44 & 1.64 & 0.37 & 0.35&0.43&& 0.05 & 0.00 & 0.95\\
25 & 0.09 & 0.38 & 0.09 & 0.09&0.10&& 0.02 & 0.00 & 0.98\\
100 & 0.05 & 0.07 & 0.02 & 0.02&0.03 && 0.00 & 0.00 & 1.00\\
270 & 0.03 & 0.01 & 0.01 & 0.01 &0.01&& 0.00 & 0.00 &1.00\\
\\
\multicolumn{5}{l}{Growth Rate}\\
10& 0.81 & 0.64 & 0.68 & 0.66 &0.68&& 0.01 & 0.99 & 0.00\\
25& 0.38 & 0.38 & 0.37 & 0.35 &0.36&& 0.00 & 0.57 & 0.43 \\
100& 0.48 & 0.47 & 0.47 & 0.46 &0.47&& 0.32 & 0.18 & 0.50\\
270& 0.38&0.40 & 0.38 & 0.38 &0.37& &0.16 & 0.05 & 0.78\\ 
 \end{tabular}
\end{center}
 \label{tabel_GDP}
\end{table}}

We then take one state at a time, and pretend state $i$ was first exposed to an intervention for in period $t\in\{T_0+1,\ldots,T\}$. If the first intervention was in period $t$, we estimate $Y_{it}(0)$ using one of the four methods using all the observations $Y_{js}$ for $s\leq t$, other than $Y_{it}$. We then compare the imputed value for $Y_{it}$ to the actual value, and square the difference. We average this over the $T-T_0$ periods where we pretend the intervention first happened. We then average this over the $N$ units to get the overall average squared error, and report the square root of this, for the four different values for $T$, and for the three different outcomes, GDP, log GDP, and the GDP growth rate, and based on horizontal or vertical cross-validation.
For the vertical crossvalidation we also report the average weights for the three individual methods.


\subsection{Discussion}

We find that the performance of the three individual methods varies widely over the settings considered here, both by the choice of outcome and by the number of time periods. Compared to the other methods the vertical / synthetic control regression does well with few periods for GDP and log GDP, but with growth rates it does relatively better with many periods. The horizontal / unconfoundedness regression does well with many periods for GDP and log GDP, but with growth rates does better for fewer periods. The factor model does poorly with the highly correlated outcomes GDP and log GDP, but does better with growth rates.

{\small
 \begin{table} \begin{center}
\begin{tabular}{lcccccccccccc}
\multicolumn{13}{c}{\sc Table II: Complexity of Models}
\\
\multicolumn{13}{c}{\sc Number of Non-zero coefficients and Ranks} \\
& \multicolumn{3}{c}{$T=10$}& \multicolumn{3}{c}{$T=25$}& \multicolumn{3}{c}{$T=100$}& \multicolumn{3}{c}{$T=270$}\\
& VT & HZ & MC& VT & HZ & MC& VT & HZ & MC& VT & HZ & MC\\\\
GDP & 3.8 & 7.5 & 1.0 &3.8 & 9.2 & 2.0 & 4.1 & 10.1 & 3.1 & 3.9 & 10.6 & 3.6 \\
Log  GDP & 16.3 & 7.5 & 1.0 &14.9 & 12.0 &1.0 &13.5 & 15.5& 1.0 & 19.0 & 21.2 & 2.0\\
Growth R & 11.8 & 3.8 & 6.9& 20.4 & 6.4 &12.3 &21.4 & 23.3&19.7&23.7 & 21.8 &32.7\\
 \end{tabular}
\end{center}
 \label{tabel_GDP2}
\end{table}}

Compare to the individual methods the
ensemble methods do well. It generally does as well or better than any of the individual methods, performing only slightly worse than the best of the individual methods in a couple of cases. If one of the individual methods does very poorly, the ensemble method takes that into account and puts little weight on that method. There is some variation by the choice of cross-validation for getting the weights, in a somewhat predictable way. With many time periods one can use the time series observations for a particular unit to choose the weights (the horizontal crossvalidation), but with few time periods one needs the cross-section variation and the vertical crossvalidation performs better.

In Table II we report the average complexity of the individual models. For the VT and HZ we report the average number of non-zero coefficients, and for the MC we report the average rank.
For GDP the vertical regression selects few states,  the horizontal regression selects relatively few lags, and the factor model selects few factors. The log GDP regression selects a substantially larger number of states in the vertical regression and more lags in the horizontal regression., and more factors. With the growth rates the horizontal model selects few lags, but the vertical model selects a substantial number of states, and the factor model is quite rich.

\section{General Discussion} 

The panel settting is an interesting one to assess model averaging and ensemble methods. There is a substantial number of conceptually very distinct methods, and  no natural model that nests them. 
We find that the relative performance of these individual methods varies widely by the relative magnitude of the number of units and time periods, as well as  by the correlation patterns in the data. None of the individual methods dominates, although the matrix completion methods typically have the advantage relative to the other methods.
However, we find that a simple linear regression approach combining the predictions based on the different methods outperforms the individual methods systematically.

\section{References}

\begin{description}

\item{Abadie, Alberto, Alexis Diamond, and Jens Hainmueller. 2010. "Synthetic control methods for comparative case studies: Estimating the effect of California’s tobacco control program." {\it Journal of the American statistical Association} 105, no. 490: 493-505.}

\item{Abadie, Alberto, Alexis Diamond, and Jens Hainmueller. 2015. "Comparative politics and the synthetic control method." {\it American Journal of Political Science} 59, no. 2: 495-510.}

\item{Athey, Susan, Mohsen Bayati, Nikolay Doudchenko, Guido Imbens, and Khashayar Khosravi. 2018. Matrix completion methods for causal panel data models. No. w25132. National Bureau of Economic Research.}


\item{Bai, Jushan, and Serena Ng. 2007. "Determining the number of primitive shocks in factor models." {\it Journal of Business \& Economic Statistics} 25, no. 1: 52-60.}

\item{Bell, Robert M., Yehuda Koren, and Chris Volinsky.  2010. "All together now: A perspective on the netflix prize." {\it Chance} 23.1: 24-29.}

\item{Breiman, Leo.  1996. "Stacked regressions." {\it Machine learning} 24.1: 49-64.}




\item{Doudchenko, Nikolay, and Guido W. Imbens. Balancing, regression, difference-in-differences and synthetic control methods: A synthesis. 2016. No. w22791. National Bureau of Economic Research.}

\item{Hoeting, Jennifer A., David Madigan, Adrian E. Raftery, and Chris T. Volinsky. 1999. "Bayesian model averaging: a tutorial." {\it Statistical science}: 382-401.}

\item{Imbens, Guido W., and Donald B. Rubin. 2015. {\it Causal inference in statistics, social, and biomedical sciences.} Cambridge University Press.}

\item{Van der Laan, Mark J., Eric C. Polley, and Alan E. Hubbard. 2007. "Super learner." {\it Statistical applications in genetics and molecular biology} 6, no. 1.}



\item{Stock, James H., and Mark W. Watson. 2006. "Forecasting with many predictors." {\it Handbook of economic forecasting} 1: 515-554.}

\item{Surowiecki, James. 2005. {\it The wisdom of crowds}. Anchor.}


\item{Varian, Hal R. 2014. "Big data: New tricks for econometrics." {\it Journal of Economic Perspectives} 28, no. 2: 3-28.}

\item{Vuong, Quang H. (1989) ``Likelihood ratio tests for model selection and non-nested hypotheses'' (1989) {\it Econometrica}: 307-333.}



\item{Zhou, Zhi-Hua. 1983.  {\it Ensemble methods: foundations and algorithms.} Chapman and Hall/CRC.}

\end{description}

\end{document}